# Indoor Sound Source Localization with Probabilistic Neural Network

Yingxiang Sun, *Student Member, IEEE*, Jiajia Chen*, Chau Yuen, *Senior Member, IEEE,* and Susanto Rahardja, *Fellow, IEEE*

*Abstract*—It is known that adverse environments such as high reverberation and low signal-to-noise ratio (SNR) pose a great challenge to indoor sound source localization. To address this challenge, in this paper, we propose a sound source localization algorithm based on probabilistic neural network, namely Generalized cross correlation Classification Algorithm (GCA). Experimental results for adverse environments with high reverberation time $T_{60}$ up to 600$m$s and low SNR such as –10dB show that, the average azimuth angle error and elevation angle error by GCA are only 4.6º and 3.1º respectively. Compared with three recently published algorithms, GCA has increased the success rate on direction of arrival estimation significantly with good robustness to environmental changes. These results show that the proposed GCA can localize accurately and robustly for diverse indoor applications where the site acoustic features can be studied prior to the localization stage.

## I. INTRODUCTION

LOCALIZATION techniques have been widely used in both outdoor environment [1] and indoor environment [2]. Diverse types of sensors including acoustic sensors, electromagnetic sensors, and optical sensors have been adopted for localization. Sensor nodes with acoustic microphones [3] with low power consumption were used in wireless sensor networks. In [4], localization on the basis of dense passive radio-frequency identification tag was proposed. Laser range finder [5] was installed on mobile robot to localize in environment where glass walls surrounded. RGB-depth camera in the light of two-dimensional light detection and ranging technique was used for localization in [6]. In contrast to common device-enable technology [7], device-free technology [8] to localize targets that do not carry any device also has appeared. Among localization techniques, indoor sound source localization (SSL) has important applications in a wide range of scenarios. For example, robots can localize the sound source to assist to detect unknown defect in smart factory. Furthermore, in smart hospital, robots can attend to patients by localizing sound source. Moreover, camera can be automatically steered for speaker localization in smart meeting room. In terms of security monitoring, robots can go on patrol and look for sound source caused by people breaking in. Therefore, indoor SSL has received a lot of attention [9] [10] in the past decades.

The existing SSL technologies can be categorized into three groups: viz, time delay estimation method, beamforming method, and machine learning method. The time delay estimation method is based on computing the time difference of arrival (TDOA). One widely used technique for TDOA is the generalized cross correlation (GCC). As reverberation and noise cause ambiguities in TDOA estimation, many efforts were made to address this problem. These works employed various types of microphone arrays, such as linear array [11], circular array [12], distributed array [13], and arbitrarily-shaped non-coplanar array [14]. The second class is the beamforming method, which can be classified into subspace approaches and beamscan approaches. Subspace approaches exploit the orthogonality between signal and noise subspaces. Two famous subspace algorithms are multiple signal classification (MUSIC) and estimation of signal parameters via rotational invariance technique (ESPRIT). Beamscan approaches can localize the array signals into one specific direction. A well-known technique is steered response power phase transform (SRP-PHAT), which is adopted by many beamscan approaches [15]-[18]. The machine learning methods are more emerging approaches and a few attempts have been made in the literature. Most of the works are supervised learning methods, including support vector machine [19], multilayer perceptron neural network [20], and Gaussian mixture model [21]. Besides, a semi-supervised learning algorithm based on manifold regularization [22] was proposed.

Although the above great works have been done to propose effective localization algorithms, there are still two more major challenges to be addressed further. The first issue is the accuracy of direction of arrival (DOA) estimation in high reverberant environments. As indoor environments are echoic, the reverberation caused by multipath propagation introduces spectral distortions and therefore severely deteriorates DOA estimation. Secondly, spectral characteristics of undesired background noise can be the same as the source signal. As such,

Manuscript received August 9, 2017; revised November 17, 2017; accepted December 6, 2017. This work was supported by the SUTD-MIT International Design Center under Grant IDG31700104 and NSFC 61750110529.

Yingxiang Sun, Jiajia Chen (Corresponding Author*) and Chau Yuen are with the Pillar of Engineering Product Development, Singapore University of Technology and Design, Singapore, (e-mail: jiajia_chen@sutd.edu.sg).

Susanto Rahardja is with School of Marine Science and Technology, Northwestern Polytechnical University, Xi'an, P.R. China, and STMIK Raharja, Tangerang, Indonesia (e-mail: susantorahardja@ieee.org).

the DOA estimation accuracy is severely degraded in low signal-to-noise-ratio (SNR) environments. Therefore, more effort is needed to improve the DOA estimation accuracy for SSL in these adverse environments. Among the applications, an important category exists where the acoustic features of the physical rooms can be pre-studied before localization. In this case, the acoustic features including the room impulse response (RIR) can be evaluated before any localization is performed, which makes machine learning methods the right tools. This kind of data driven training methods can be more effective especially when the environment is too complex to be modeled.

In this paper, we propose a probabilistic neural network (PNN) based SSL algorithm for the applications where pre-localization site survey is possible. Compared with other existing machine learning methods, the most important advantage of PNN is that it does not require any iterative training. In addition, the GCC feature is adopted to robustly represent the sound source position, making the training procedure effective in reverberant and noisy environments. Finally, the proposed weighted location decision method improves the accuracy of the DOA estimation by revisiting and accessing the probabilities of the adjacent clusters. Owing to these novelties, the results show that the proposed algorithm can perform more accurate SSL than existing methods in the adverse environments. The performance is proven to be robust too, when room environment and/or geometry varies.

## II. SYSTEM MODEL AND PROBLEM FORMULATION

In this section, we present the SSL problem to be addressed. We consider the problem of stationary single source localization inside a 3-dimensional rectangular enclosed room. The location of the source is arbitrary inside the room. A stationary microphone array which consists of $M$ microphones is used to receive sound signals inside the same room. Through these fixed microphones in the array, we can receive the signal transmitted from the source directly and the delayed replicas of the source reflected by room surfaces. The $m^{th}$ microphone can be represented as $M_m$ with $m \in [1, M]$. When the sound wave hits a surface such as a wall, a floor or a ceiling, part of the wave is absorbed by the surface while the rest is reflected back into the room. We assume that the sound wave is reflected by the surfaces with the angle of incidence equal to the angle of reflection. Therefore, the received signal at each microphone is a mixed signal, consisting of the signal transmitted from the source directly and the delayed replicas of the source which are reflected and attenuated.

If the source signal is $s(t)$, the received signal $x_m(t)$ at the $m^{th}$ microphone can be expressed as

$$x_m(t) = h_m(t) \otimes s(t) + n_m(t), \quad (1)$$

where $\otimes$ denotes convolution. $n_m(t)$ is the noise at the $m^{th}$ microphone, which is uncorrelated with $s(t)$ and those noises at other microphones [9]. $h_m(t)$ is the RIR which contains the multipath propagation and attenuation information between the sound source and the $m^{th}$ microphone. $h_m(t)$ varies with sound source and the $m^{th}$ microphone positions. By assuming the received signals set $X = [x_1, x_2, \ldots, x_M]^T$, RIR set $H = [h_1, h_2, \ldots, h_M]^T$ and noise set $N = [n_1, n_2, \ldots, n_M]^T$, (1) can be written as

$$X = H \otimes s(t) + N. \quad (2)$$

If we divide a room into a set of space clusters whose volumes are small enough, each space cluster can be represented by a unique 3-dimensional coordinate inside it. To cope with the high computational burden, the regressive SSL inside a 3-dimensional room can be transformed into a likelihood based nonlinear classification problem. Therefore, the classifier can decide which particular cluster the source belongs to, as shown in Fig. 1. In this classification problem, each space cluster is a category and a total number of $K$ categories can be created, i.e. $C = [c_1, c_2, \ldots, c_K]$ and $c_i \in R^3$ with $i \in [1, K]$. The complexity of the classification grows with the increase of $K$ for finer-grained clusters, which leads to a more accurate localization if the classification is successful. All the $K$ categories are possible solutions and each possible solution $c_i$ has a set of features $feature_i$ that decide the probability of $c_i$ being the final solution. Based on the features and the received signals set $X$, a dedicated classifier classifies the source into one cluster $c_s$, whose unique coordinate representative is $[d_{x,s}, d_{y,s}, d_{z,s}]$. $c_s$ is the solution of the classification problem while $[d_{x,s}, d_{y,s}, d_{z,s}]$ is the solution of SSL problem. This classification problem by classifier function **classify**($\cdot$) can be summarized as

$$c_s = \textbf{classify}(X, \sum_{\forall i} feature_i). \quad (3)$$

Assume the actual source location is $[s_x, s_y, s_z]$ inside the cluster $c_{source}$. Even the classification solution is wrong if $c_s \neq c_{source}$, the regression localization error $\varepsilon$ can be evaluated as

$$\varepsilon = \sqrt{(d_{x,s} - s_x)^2 + (d_{y,s} - s_y)^2 + (d_{z,s} - s_z)^2}. \quad (4)$$

The DOA results in terms of $\theta$ and $\phi$ can be obtained from

$$d_{x,s} = x_m + r\sin\theta\cos\phi, d_{y,s} = y_m + r\sin\theta\sin\phi, d_{z,s} = z_m + r\cos\theta, \quad (5)$$

where $x_m$, $y_m$ and $z_m$ are coordinates of the microphone array. $r$ denotes the distance between the cluster $c_s$ and array center. $\theta \in [-90°, +90°]$ is the elevation angle, from $r$'s orthogonal projection onto the $xy$-plane towards the positive $z$-axis. $\phi \in (-180°, +180°]$ is the azimuth angle, from the positive $x$-axis towards the positive $y$-axis, in terms of $r$'s orthogonal projection onto the $xy$-plane.

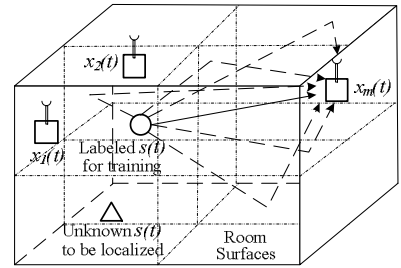

Fig. 1. Space cluster classification for SSL

If the longest diagonal inside one cluster is $l$, $\varepsilon$ is bounded by $l$ when the classification is correct. If the classification is incorrect but $c_s$ is an adjacent cluster of $c_{source}$, it is still possible to have $\varepsilon$ bounded by $l$. Therefore, the localization error bounding depends on the correctness probability of the classification as

$$P(\varepsilon \leq l | c_s = c_{source}) = 1; \; P(\varepsilon \leq l) \geq P(c_s = c_{source}). \quad (6)$$

To minimize $\varepsilon$, therefore, we need an efficient and accurate classifier which is with high classification correctness rate and affordable computational complexity. In the next section, we will present the details of the proposed SSL algorithm based on the PNN classifier.

### III. THE PROPOSED ALGORITHM

The relationship between the source position and the recorded signals at microphone array is nonlinear. We adopt PNN [23] as the classifier, because PNN is more suitable for the nonlinear multi-classification problem. PNN contains four layers, i.e. input layer, pattern layer, summation layer and decision layer successively. With this classifier, we propose a GCC classification algorithm (GCA) to solve the classification problem formulated in Section II.

#### A. GCC Feature Extraction

In order to generate the input vector space $I$ for the PNN, we need to extract features from the signals at microphone array. The feature of each received signal is unique. Meanwhile, as each sound source is located at a unique position, there is a one-to-one correlation between the received signals and source positions. For a machine learning algorithm to provide good solution, it is essential to select well-defined features prudently for the training. The reason is that the probability densities of the category patterns are unknown initially. The derivation of these probability densities solely relies on these selected features. GCC is an ideal candidate to be used as feature, since it contains all the needed information for DOA estimation and is reliable in reverberant and noisy environments.

GCC varies across different frames. Taking the silent frame as an example, GCC is mainly due to the noise. In this case, if we directly use GCC in a single frame as the feature, it is not representative. Therefore, GCC from higher SNR frames need to be evaluated with higher weightage, while the rest are with lower weightages or even neglected. In our method, GCC from all frames are weighted and summed to be the feature [20], namely GCC feature. As GCC and the weights for each frame are different, GCC feature is unique. The length $L$ of each frame is selected based on compromise between good spectral resolution and small bias and variance. Therefore, a vector consisting of $L$ GCCs can be extracted for each frame. Assume that the source signal consists of totally $F$ frames. We use $GCC_f^l$ to represent the $l^{th}$ GCC corresponding to the $f^{th}$ frame of the source signal, with $l \in [1, L]$ and $f \in [1, F]$. The GCC feature corresponding to one sound source can be expressed as

$$GCC = \sum_{f=1}^{F} w_f \cdot \sum_{l=1}^{L} GCC_f^l, \quad (7)$$

where

$$w_f = \frac{\sum_{l=1}^{L} |GCC_f^l|^\gamma}{\sum_{f=1}^{F} \sum_{l=1}^{L} |GCC_f^l|^\gamma}, \quad (8)$$

denoting the weight of the $f^{th}$ frame. $\gamma$ is a tuning parameter.

To localize a sound source by an $M$-microphone array with $M \geq 2$, we can compute a total number of $M(M-1)/2$ GCC features using (7), with each corresponding to one microphone combination. These $M(M-1)/2$ GCC features are grouped together to form the complete GCC set corresponding to the sound source. Therefore, more accurate SSL can be achieved with more microphone combinations, but at the expense of higher computational complexity in GCC feature extraction.

#### B. Training

At the beginning of the training, the enclosed room is divided into a number of $K$ equal-dimension rectangular clusters, namely $c_1, c_2, \ldots, c_i, \ldots, c_K$ with $i \in [1, K]$. This dividing procedure is defined as **cluster**($Dim, K$), where the dimension of each cluster $Dim$ depends on the required localization accuracy. Assume that $n_i$ is the total number of training samples taken inside the $i^{th}$ cluster, we can define four vector spaces, namely $X=\{X_{i,j}\}$, $S=\{S_{i,j}\}$, $GCC=\{GCC_{i,j}\}$ and $H=\{H_{i,j}\}$. Each $X_{i,j}$ represents signals produced at the microphone array $M$ when the $j^{th}$ training sample sound source $S_{i,j}$ inside the $i^{th}$ cluster is placed, with $j \in [1, n_i]$. $GCC_{i,j}$ is the corresponding GCC feature extracted from $X_{i,j}$. $H_{i,j}$ represents the corresponding RIR between $M$ and $S_{i,j}$.

Given the sampling frequency of sound signal ($f_{sample}$), the absorption coefficient of the room ($\alpha_c$), sound velocity in the air ($v_c$), reverberation time ($T_{60}$) and the noise in the room ($N$), the RIRs between the microphone array and sources can be computed [24]. This procedure is defined as **RIR**($f_{sample}, v_c, T_{60}, N, \alpha_c, M, S$). By convoluting $H$ with $S$ and adding $N$, we can produce the signal vector space $X$. After that, the GCC features $GCC$ are extracted using (7). We define this procedure as **GF**($X, \gamma$). Upon completion of the feature extraction, all features are supplied to PNN as the input vector space $I$. The number of neurons of input layer is equal to the dimension of input GCC feature vector. In pattern layer, the number of neurons equals to the total number of training samples placed to train the PNN. Therefore, there are $\sum_{i=1}^{K} n_i$ neurons in pattern layer. The neurons of the pattern layer map input GCC feature vector to a high-dimensional space and estimate corresponding probabilistic density by Gaussian kernel represented as

$$\varphi_{i,j}(GCC) = \frac{1}{(2\pi)^{D/2} \sigma^D} \exp\left[-\frac{(GCC - GCC_{i,j})^T (GCC - GCC_{i,j})}{2\sigma^2}\right], \quad (9)$$

where $\varphi_{i,j}(GCC)$ is the Gaussian kernel function. $\sigma$ is the spread parameter which represents the width of the Gaussian kernel. $T$ denotes the transpose. $GCC$ is the $D$-dimensional input GCC feature vector. $GCC_{i,j}$ is the center of the kernel.

The output of each neuron in the pattern layer can be generated using (9) and all outputs are transmitted to the summation layer, in which the number of neurons equals $K$. By averaging the output of all neurons that belong to the same cluster $c_i$, the summation layer computes the probability $p_i(GCC)$ of that input GCC feature being classified into the $i^{th}$ cluster as

$$p_i(GCC) = \frac{1}{(2\pi)^{D/2} \sigma^D} \frac{1}{n_i} \sum_{j=1}^{n_i} \exp\left[-\frac{(GCC - GCC_{i,j})^T (GCC - GCC_{i,j})}{2\sigma^2}\right]. \quad (10)$$

Assume the priori probability of occurrence of every cluster $c_i$ is $h_i$, and the loss caused by misclassification decision for each cluster $c_i$ is $co_i$. The decision layer neuron classifies the input GCC feature into cluster $c_s$ according to the Bayes's decision rule [23] as

$$h_s \times co_s \times p_s(GCC) > h_i \times co_i \times p_i(GCC), \quad \forall i \neq s, \quad (11)$$

where $p_s(GCC)$ is the probability of $GCC$ being classified into cluster $c_s$.

We assume $h_i$ and $co_i$ are unique for all the clusters, so that the GCC feature is classified into cluster $c_s$ as

$$c_s = \arg\max \{p_i(GCC), c_i\}. \quad (12)$$

We define this procedure as **DA(*GCC*)**. $p_i(GCC)$ also is the probability of each training sample being classified into the $i^{th}$ cluster, as there is a one-to-one correlation between *GCC* and *S*. In terms of the output layer, there is only 1 neuron, as only the most probable class is chosen by the PNN.

### C. Localization

Once the PNN is trained with the GCC features, the GCA continues to the second stage to localize the unknown sound source $S_u$ into one of the *K* clusters. As presented in Section III B, the probability of $S_u$ being classified into every cluster can be computed by PNN using (10), according to $S_u$'s GCC feature.

Therefore, the decision layer classifies $S_u$ into any of the *K* clusters $c_s$ using (12) with those computed probabilities. However, when the space cluster's volume is small, it is difficult to distinguish which cluster the source actually belongs to and hence the rate of misclassification becomes higher. The situation gets worse when the actual source $c_{source}$ is close to the boundaries of two adjacent clusters. To solve this problem, we propose a weighted location decision method (WLDM) in GCA instead of using the PNN decision layer to classify directly, which is presented below.

To guarantee $\sum_{a=1}^{K} p_a = 1$, the softmax function is adopted to be the transfer function between the pattern layer and the summation layer. Thereby we can normalize the categorical probability distribution in the range of (0, 1) that adds up to 1. With the probabilities of all clusters computed, we select the $\zeta$ most possible clusters whose probability sum is less than a cluster size dependent on a threshold *THR*, i.e. $\sum_{a=1}^{\zeta} p_a \leq THR$. The selection starts from the cluster with top probability following the descending order, and stops before one additional cluster that will cause the probability sum to be higher than the threshold. After these $\zeta$ adjacent clusters are selected, we perform the localization through the following two steps, which are preliminary estimation and sample points estimation.

Let $P_a$ denote the central point chosen for the $a^{th}$ cluster, with $a \in [1, \zeta]$ and its Cartesian coordinates are $x_a$, $y_a$ and $z_a$. The preliminarily estimated source position $P_s$ with Cartesian coordinates $x_s$, $y_s$ and $z_s$ are computed as

$$x_s = \sum_{a=1}^{\zeta} p_a \cdot x_a;\ y_s = \sum_{a=1}^{\zeta} p_a \cdot y_a;\ z_s = \sum_{a=1}^{\zeta} p_a \cdot z_a. \quad (13)$$

This procedure is defined as **PE**($p_a$, $P_a$). With (13), we can compute the distance $l_a$ between the representative point of the $a^{th}$ adjacent cluster and the estimated source position by

$$l_a = \sqrt{(x_a - x_s)^2 + (y_a - y_s)^2 + (z_a - z_s)^2}. \quad (14)$$

The longer distance indicates that the actual source position is more likely to be far away from that particular cluster and hence its probability is supposed to be reduced. Therefore, new weight of the $a^{th}$ cluster which is inversely proportional to the distance can be derived by

$$w_a = \frac{(1/l_a)^{\lambda}}{\sum_{a=1}^{\zeta} (1/l_a)^{\lambda}}, \quad (15)$$

where $w_a$ is the new weight of the $a^{th}$ cluster. $0<\lambda<1$ denotes the controlling parameter. This procedure is defined as **weight$_{cluster}$**($l_a$, $\lambda$).

In order to reduce the error further, we adjust the localization by more sample points in the second step. In each adjacent cluster, $\beta$ sample points are selected to represent the cluster position more accurately. Similar to the new weights of cluster, $\beta$ sample point weights can be computed by

$$w_{a,t} = \frac{(1/l_{a,t})^{\rho}}{\sum_{t=1}^{\beta} (1/l_{a,t})^{\rho}}, \quad (16)$$

where $l_{a,t}$ is the distance from $P_s$ to the $t^{th}$ sample point in the $a^{th}$ cluster with $t \in [1, \beta]$. $w_{a,t}$ denotes the weight for the $t^{th}$ sample point in the $a^{th}$ cluster. $0<\rho<1$ is the controlling parameter. This procedure is defined as **weight$_{sp}$**($l_{a,t}$, $\rho$). Therefore, we can decide the localization of $c_s$ through **WLDM**($w_a$, $w_{a,t}$, $P_{a,t}$):

$$d_{x,s} = \sum_{a=1}^{\zeta} w_a \cdot \left( \sum_{t=1}^{\beta} w_{a,t} \cdot x_{a,t} \right);\ d_{y,s} = \sum_{a=1}^{\zeta} w_a \cdot \left( \sum_{t=1}^{\beta} w_{a,t} \cdot y_{a,t} \right);$$
$$d_{z,s} = \sum_{a=1}^{\zeta} w_a \cdot \left( \sum_{t=1}^{\beta} w_{a,t} \cdot z_{a,t} \right), \quad (17)$$

where $x_{a,t}$, $y_{a,t}$ and $z_{a,t}$ are Cartesian coordinates of the $t^{th}$ sample point in the $a^{th}$ cluster.

TABLE I  THE PSEUDO CODE OF THE PROPOSED GCA

```
GCA(Train, Localize)
begin
  Train(M, Dim, f_sample, v_c, T_60, N, α_c, K, n_i, γ, S) // training stage of GCA
  begin
    C=cluster(Dim, K); // divide the room into K clusters
    for all i ∈ [1, K]
      for all j ∈ [1, n_i]
        H_{i,j}=RIR(f_sample, v_c, T_60, N, α_c, M, S_{i,j}); // compute the RIR
        X_{i,j}=H_{i,j}⊗S_{i,j}+N; // obtain the signal at microphone array
        GCC_{i,j}=GF(X_{i,j}, γ); // extract GCC feature
      end
    end
    p_i(GCC)=DA(GCC); // train PNN
  end
  Localize(S_u, γ, ζ, β, λ, ρ, THR); // localization stage of GCA
  begin
    GCC=GF(S_u, γ); p_i(GCC)=DA(GCC); // compute the probability
    for all a ∈ [1, ζ]
      P_s=PE(p_a, P_a); // obtain preliminary estimation of source position
      w_a=weight_cluster(l_a, λ); // compute weights of clusters
      for all t ∈ [1, β]
        w_{a,t}=weight_sp(l_{a,t}, ρ); // compute weights of sample points
      end
    end
    c_s=WLDM(w_a, w_{a,t}, P_{a,t}); // obtain final source position
  end
  return DOA=[θ, φ];
end
```

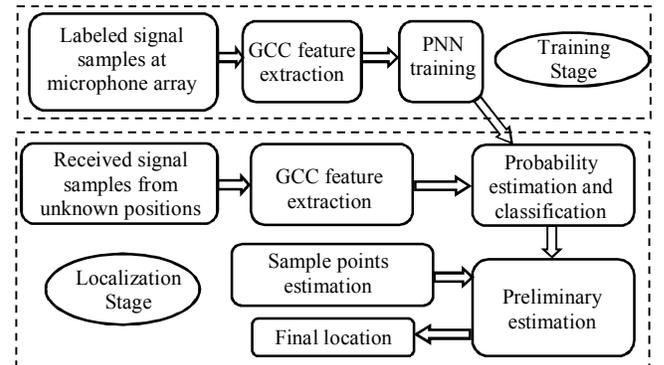

Fig. 2. Flow chart of the proposed GCA

The located position $c_s=[d_{x,s}, d_{y,s}, d_{z,s}]$ is the solution of the GCA in Cartesian coordinates and $[\theta, \phi]$ is the DOA results which can be computed by (5). The pseudo code of the proposed GCA is summarized in Table I. The function **GCA**(*Train*, *Localize*) consists of two sub-functions which are **Train**($M$, $Dim$, $f_{sample}$, $v_c$, $T_{60}$, $N$, $\alpha_c$, $K$, $n_i$, $\gamma$, $S$) and **Localize**($S_u$, $\gamma$, $\zeta$, $\beta$, $\lambda$, $\rho$, $THR$), representing the two stages of GCA respectively. Finally, the DOA ($\theta, \phi$) is returned as the outputs. The flow chart of the proposed GCA is depicted in Fig. 2.

## IV. SYNTHETIC EXPERIMENTAL RESULTS AND DISCUSSION

In this section, synthetic experiments are conducted to evaluate the performance of the proposed GCA while other three recently published algorithms presented in [14], [18], and [19] are employed to be the competing methods.

### A. Synthetic Experimental Setup

A typical medium size meeting room with dimension as 4.0m × 4.0m × 4.0m is simulated. The microphone array consists of six microphones, which are placed at $M_1$=(1.8m, 2.0m, 2.0m), $M_2$=(2.2m, 2.0m, 2.0m), $M_3$=(2.0m, 1.8m, 2.0m), $M_4$=(2.0m, 2.2m, 2.0m), $M_5$=(2.0m, 2.0m, 1.8m) and $M_6$=(2.0m, 2.0m, 2.2m). The source is placed on a sphere centered at the centroid of the room, with three different radius values 0.5m, 1.0m and 1.5m. On each of the three spherical surfaces, the sound source is placed at 21 different azimuth values from −160º to +160º and at 9 different elevation values from –60º to +60º, both with even intervals. In total, the sound source is placed at 567 different positions distributed in the room. In our experiments, omnidirectional microphones are adopted, with frequency response from 20Hz to 20$k$Hz and dynamic range of 87dB.

We use six microphones rather than other numbers to form the array with such spatial distribution mainly due to four reasons. Firstly, if microphones are distributed along each dimension of the space, position of the source can be better determined as the sound propagates via each dimension of room. Secondly, we only use 2 microphones along each of the three dimensions to minimize computational complexity. Thirdly, considering the tradeoff between computational complexity and validness of information obtained from cross correlations, we set the maximum distance between any two microphones to be 40cm. In addition, considering a test source can be placed anywhere in the room, by referring to the setup of competing method TDE [14], the center of microphone array is placed at the center of the room.

A clean speech sampled at 8$k$Hz as [25] is adopted to be the sound source. The 2.7-second speech (from 220Hz to 3.4$k$Hz) is from the NOIZEUS database in American English language. The sound source is also omnidirectional in the setup. The reverberation time $T_{60}$, which measures the time for the original sound to decay by 60dB, is set to be different levels as 0$m$s, 100$m$s, 200$m$s, 400$m$s and 600$m$s. The longer $T_{60}$ represents the higher reverberation in the room. The SNR in the room is set to be different levels as 10dB, 0dB, –5dB and –10dB, where the noise is additive noise. The duration of each frame of the speech signal is chosen to be 0.064s and the overlap rate between two frames is set to be 62.5%. As the maximum distance between any two microphones of the array is 0.4m, the maximum possible time delay is 1.17$m$s by assuming the sound speed in the air being 343m/s. As the sampling rate is 8$k$Hz, the maximum delay number in samples is 10. Therefore, for a microphone pair, the first 10 cross correlations contain the valid information. However, in case of missing validity, we select the first 16 cross correlations to be the feature. As there are totally 15 microphone combinations for cross correlation computing, the dimension of the GCC feature vector applied to the input layer is 240. Therefore, the input layer consists of 240 neurons.

In the training stage of our synthetic experiments, the room is divided into 4096 equal-dimension rectangular space clusters with dimensions as 0.25m × 0.25m × 0.25m each. The sound source is randomly and successively placed in each cluster only once, i.e. $n_i$=1, as the cluster volume is small. Therefore, a total of 4096 complete GCC feature sets are extracted. In this case, both pattern layer and summation layer consist of 4096 neurons. For the spread parameter $\sigma$, a small $\sigma$ will cause overfitting while a large $\sigma$ will result in underfitting. In practice, by referring to [23], $\sigma$ can be selected from 3 to 10. In our experiments, we set its value to be 5. For the WLDM, we select the 15 most possible clusters whose probability sum is less than 0.004. For the controlling parameters, both $\lambda$ and $\rho$ are set to be 0.25 while $\gamma$ is set to be 2. In each adjacent cluster, 8 vertexes are selected as sample points to represent the cluster position more accurately.

### B. Implementation

We perform the synthetic experiments to compare our results with three recent methods, which are time delay estimation method TDE [14], beamforming method TL-SSC [18], and machine learning method LS-SVM [19]. In the synthetic experiments, $Dim$, $T_{60}$, $\alpha_c$, and SNR are all required by our proposed method and the competing methods. As the author-shared codes of TDE and TL-SSC are available online at [26] and [27] respectively, we select these two algorithms as competing methods. This helps to avoid any potential errors when modeling the algorithm by non-authors so that the comparison is fair and valid. As the TL-SSC is an improved version of the widely used SRP-PHAT algorithm, we do not adopt the original SRP-PHAT algorithm as a competing method. For LS-SVM, we collect the TDOA features as its original paper [19] for training. In addition, as LS-SVM algorithm transforms localization to be a pure classification problem, we assume that the estimated sound source position is at the centroid of the cluster where it is classified into. Furthermore, the performances of these competing methods degrade if we adopt our microphone array setup into their methods. To make fair comparison, therefore, the microphone arrays for these three competing methods are setup in the same way as given in their original papers, [14], [18], and [19] respectively. What's more, to improve RIR computation efficiency, fast image method [24] is adopted and the source code is available online at [28]. All the four methods are implemented in Matlab and run by a workstation with 32GB RAM and dual Intel Xeon 2.4GHz processor E5-2630 V3.

### C. Results and Discussion

#### Validation on Feature Extraction

The first experiment is to examine the effectiveness of GCC features. Simulations are performed in four different environments with SNR decreasing from 10dB to –10dB, when $T_{60} = 0ms$, as demonstrated by (a) to (d) of Fig. 3. With the same acoustic environment as in each subplot, we repeated the extractions for three times which are separated by the two black lines. It can be seen that the computed GCC features in yellow color patterns demonstrate good representativeness for the testing clusters. The contrast between the yellow patterns and the blue regions becomes more distinctive when SNR rises from –10dB to 10dB. This shows that the GCC feature representativeness is more reliable when SNR is high. Similar regularity can be observed when SNR is fixed and $T_{60}$ varies, where GCC feature is more reliable when $T_{60}$ is low.

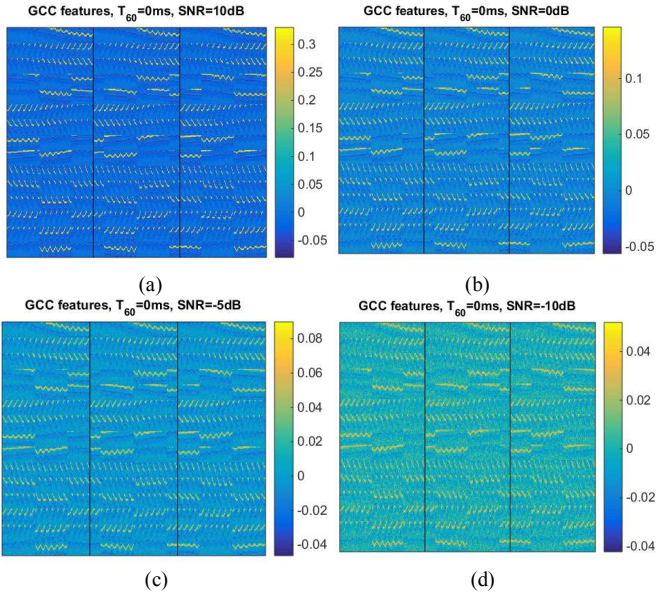

Fig. 3. GCC features extracted from different SNR when $T_{60}=0ms$

### Impact of Reverberation and SNR

With the validated GCC features, we perform the SSL using GCA and compare the performance. Because most of the SSL applications are to localize the source directions, we adopted the DOA as the performance metric. The results are summarized in Table II. A total of 20 different acoustic environments are created by varying SNR from 10dB to –10dB and $T_{60}$ from 0ms to 600ms. With each environment, the DOA estimation error (DEE) in terms of the mean error and standard deviation of $\phi$ and $\theta$ are collected from the 567 localized positions. To evaluate the performance with the different accuracy requirements, the localization successful rate for DOA estimation (SRDE) is defined. SRDE($\alpha$º) presents the percentage of localizations with both $\phi$ error and $\theta$ error less and equal to $\pm \alpha$º out of the 567 localizing.

In Table II, the results of SRDE(10º), SRDE(20º) and SRDE(30º) are provided. When SNR is fixed and $T_{60}$ varies from 0ms to 600ms, the accuracy generally drops for every algorithm. Similar trend can be observed when $T_{60}$ is fixed and SNR decreases. However, SRDE is not always increasing with the increased SNR. In some scenarios, such as very long $T_{60}$, SRDEs may not strictly increase with the SNR. This shows that when reverberation is severe, a little vary of SNR will not affect the SRDE significantly. If we compare across different algorithms, the proposed GCA outperforms other three algorithms significantly. SRDE(30º) of GCA is 100% when $T_{60}$ is low, regardless of SNR, and drops to 69.8% in the worst case of $T_{60} = 600ms$ and SNR = –10dB.

For TDE and TL-SSC, SRDE(30º) achieves 54.9% and 64.2% in the best case of $T_{60} = 0ms$ and SNR = 10dB. With adverse environments, however, the SRDE of TDE and TL-SSC drops, which shows that high reverberation and low SNR affect the localization effectiveness of these two algorithms. When $\alpha$º is small such as 10º, the baseline successful rate by random localization should be (20º/360º) × (20º/180º) = 0.62%. In the most adverse environment, TDE provides low SRDE(10º) slightly better than this baseline rate. Nevertheless, this can still reasonably show that TDE performance will drop for more adverse environments.

For LS-SVM, the results for $T_{60} = 600ms$ are left blank in Table II, as the provided source code of fast image method encounters errors in this case. To avoid inappropriate implementation of LS-SVM, we present and discuss the results of LS-SVM when $T_{60}$ varies from 0ms to 400ms only. The results show that LS-SVM has performance similarly to TDE and TL-SSC in terms of accuracy for low reverberation and high SNR but drops in very adverse environments.

When SNR = –10dB, the SRDEs by the competing algorithms with longer $T_{60}$ are sometimes slightly higher than those with shorter $T_{60}$. It shows that these algorithms are more sensitive to the extremely low SNR. When the signals are very weak, the algorithms are significantly affected by the noises.

The averages of DEEs and SRDEs under the twenty different environments are computed for each algorithm. They are plotted in Fig. 4(a) and (b) respectively. In Fig. 4(a), the average of mean errors of azimuth angle and elevation angle by GCA are only 4.6º and 3.1º respectively, indicating that it can estimate DOA very accurately. In contrast, the DEEs of other algorithms are significantly higher. Comparing with the best performance among the three competing algorithms, GCA can localize with average of 88.6% and 83.8% reduced $\phi$ error and $\theta$ error respectively, for all the 20 acoustic environments. In Fig. 4(b), the average SRDE(10º), SRDE(20º) and SRDE(30º) by GCA can achieve 87.5%, 94.4% and 96.9% respectively. On the other hand, the averages of SRDEs of other three methods in the 20 different acoustic environments are significantly lower. Compared with the best performances among the three algorithms, GCA improves averages of SRDE(10º), SRDE(20º) and SRDE(30º) by 81.1%, 74.1% and 60.3% respectively.

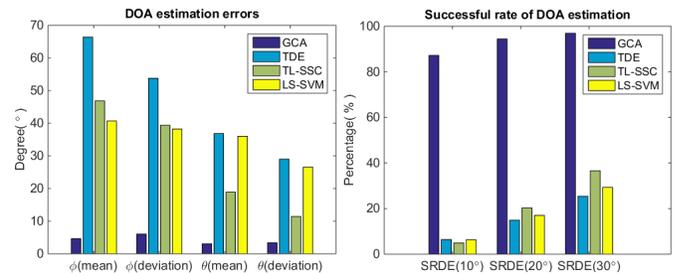

Fig. 4. (a) DOA estimation errors (b) SRDEs with different requirements

However, it should be noted that the significantly increased SRDEs by GCA are mainly contributed by several factors. The first one is the pre-localization site survey effort to collect the features, which is not needed by TDE and TL-SSC. Secondly, the 567 different positions tested in this experiment covered most of the positions in the room. Because the performance of other competing algorithms may vary when testing at different positions, the consistent performance of GCA becomes significant when comparing the average SRDEs of 567 tests. What's more, for LS-SVM, when the space cluster's volume is small, it is difficult to decide which cluster the sound source belongs to. This is a reason why LS-SVM cannot achieve even higher SSL accuracy, meanwhile, it verifies the contribution by the proposed WLDM in GCA, which can solve this problem. In addition, the GCC features used in GCA is more robust compared to TDOA features used in LS-SVM.

*Robustness Validation*

In practice, when room geometry and acoustic features change, such as people movement, doors opening and closing, the validity of the collected training data varies. In this case, we need to ensure that the DOA by GCA is still accurate and robust even when the environment changes after training. Moreover, we also expect that the proposed GCA can perform well with sound sources at different frequencies.

For the first experiment, we evaluate the robustness of GCA with respect to the change of reverberation time. We collect four groups of training data with SNR = –10dB, –5dB, 0dB and 10dB respectively, when $T_{60}$ = 200$m$s. Next, for each group, we vary $T_{60}$ to 0$m$s, 100$m$s, 400$m$s and 600$m$s to reflect the actually changed $T_{60}$ during localizing and collect the testing data from the 567 test positions. The results are summarized in Table III(A). The $T_{60}$ = 200$m$s results collected as training data is the benchmark and highlighted in grey color. Compared to the benchmarks, the worst SRDE(30º) drops are 22.7%, 6.2%, 1.1% and 0.5% respectively for the four groups. This shows that GCA is robust to $T_{60}$ with SRDE(30º) even when $T_{60}$ varies significantly, except the very adverse environment where SNR = –10dB and $T_{60}$ = 600$m$s.

For the second experiment, we validate the robustness of GCA with respect to the change of SNR. Similarly, we collect five groups of results with different $T_{60}$ as 0$m$s, 100$m$s, 200$m$s, 400$m$s and 600$m$s respectively when SNR = 0dB, as the training data. Next, we vary SNR to –10dB, –5dB and 10dB to reflect the actually changed SNR during localizing and collect the testing data from the 567 test positions. The results are summarized in Table III(B). The results with SNR = 0dB is the benchmark and highlighted in grey color. Compared to the benchmarks, the worst SRDE(30º) drops are 5.7%, 6.7%, 5.1%, 29.4% and 36.2% respectively for the five groups. Therefore, GCA is robust to SNR except the very adverse environments where SNR = –10dB meanwhile $T_{60}$ = 400$m$s and 600$m$s.

For the third experiment, we illustrate the impact of frequency change on localization accuracy of GCA. We use two new sound sources [29] rather than human speech, i.e. machinery sound and telephone ring, whose frequencies are different from the preceding human speech source. We conduct the experiment under the conditions where SNR = –10dB and 10dB, with $T_{60}$ varying from 0$m$s to 600$m$s. All the setups are the same as those of human speech scenario. The results are summarized in Table III(C). Compared to the results of human speech scenario in Table II, when $T_{60}$ = 0$m$s, 100$m$s, and 200$m$s, SRDE(30º) is almost unchanged for both of the two new sources. When $T_{60}$ becomes higher with SNR = –10dB, SRDE(30º) of machinery sound increases a little while that of telephone ring decreases slightly. When $T_{60}$ becomes higher with SNR = 10dB, SRDEs(30º) of both machinery sound and telephone ring drop slightly faster. However, on average 94.9% accuracy and 88.7% accuracy can still be achieved in terms of SRDE(30º) for these two new sound sources respectively. Therefore, we can conclude that, the localization accuracy of the proposed method is slightly affected by the changes of frequency, and therefore good performance still can be achieved when frequency changes.

*Impact by Different Test Set*

To evaluate the performances of the four algorithms with different test set, we re-compute the SRDEs of the four algorithms with 378 positions. To make sure we evaluated the competing algorithms in the correct manner, we have used the author-shared source codes of TDE and TL-SSC. These positions are obtained by removing the source positions on the spherical surface with radius = 1.0m, from the previous 567 positions. During the experiment, the SNR is set to be –10dB, –5dB, 0dB, and 10dB, while $T_{60}$ varies from 0$m$s and 600$m$s. The results are summarized in Table IV. It can be observed that the three competing algorithms can perform relatively well when SNR is higher, i.e. SNR = 0dB and 10dB, compared with the cases where SNR is lower, i.e. SNR = –10dB and –5dB. However, GCA still outperforms them even with higher SNR. This shows that GCA performs better than other three algorithms when the sound source is either close to or far away from the microphone array in adverse environments. This better performance results from several factors, such as the acoustic feature studying, robust GCC feature, the proposed WLDM method, and consistent localization capability at different test positions in a room.

*Complexity Comparisons*

The computational complexities of the four algorithms at 567 positions are summarized in Table V. For each $T_{60}$, the presented CPU time and real processing time are the average results when SNR = –10dB, –5dB, 0dB and 10dB. As more than one core are called during computing, the CPU time is higher than the real processing time. From Table V, the machine learning based GCA and LS-SVM have dominating offline training, which consists of RIR computing, feature extraction and training. However, we can overcome this defect by taking further approximation with fast image method and using less training samples. In addition, offline training in the machine learning algorithm is only performed once for the fixed room. For GCA, the CPU time of the rest online localization only accounts for 0.88%, 0.55%, 0.32%, 0.13%, and 0.06% of the total CPU time for T60 = 0$m$s, 100$m$s, 200$m$s, 400$m$s, 600$m$s respectively. This makes GCA especially suitable for the real-time localization applications when pre-localization site survey has already been done. In contrast, LS-SVM is computationally inefficient when the number of classification categories is large and the offline training costs more than 367334.3 seconds of CPU time. This validates the advantage of GCA on training speed compared with LS-SVM. TDE costs at least 32911.0 seconds of CPU time to generate the results of localizing 567 positions. In contrast, TL-SSC is the most computationally efficient algorithm among the four methods. The pre-localization look-up table (LUT) computing costs about 906.9 seconds CPU time and the localization costs around 755.4 seconds only. Compared with the time cost of

TABLE II  RESULTS OF THE FOUR ALGORITHMS AT 567 TEST POSITIONS

| SNR/dB | | −10 | | | | | −5 | | | | |
|---|---|---|---|---|---|---|---|---|---|---|---|
| $T_{60}$/ms | | 0 | 100 | 200 | 400 | 600 | 0 | 100 | 200 | 400 | 600 |
| **GCA (Proposed)** | $\phi$ error (mean/deviation)/° | 1.8/1.6 | 2.4/2.0 | 5.1/5.0 | 12.7/14.0 | 22.8/28.2 | 1.2/0.9 | 1.3/1.1 | 2.1/1.8 | 5.4/6.6 | 10.8/15.3 |
| | $\theta$ error (mean/deviation)/° | 1.3/1.1 | 1.7/1.3 | 3.6/3.2 | 8.7/9.3 | 14.3/15.0 | 0.9/0.7 | 1.2/0.9 | 1.5/1.2 | 3.6/3.9 | 7.0/8.3 |
| | SRDE(10°) | 100% | 99.6% | 84.3% | 46.7% | 27.7% | 100% | 100% | 99.5% | 81.8% | 56.3% |
| | SRDE(20°) | 100% | 100% | 97.9% | 75.1% | 55.2% | 100% | 100% | 100% | 96.0% | 82.7% |
| | SRDE(30°) | 100% | 100% | 99.8% | 88.7% | 69.8% | 100% | 100% | 100% | 98.4% | 90.8% |
| **TDE[14]** | $\phi$ error (mean/deviation)/° | 87.2/53.1 | 86.9/54.1 | 82.4/54.8 | 83.2/51.5 | 86.7/54.0 | 72.8/57.3 | 68.4/54.2 | 66.7/53.0 | 76.2/55.4 | 79.7/56.7 |
| | $\theta$ error (mean/deviation)/° | 43.7/32.9 | 44.7/32.1 | 43.3/31.5 | 42.8/30.4 | 43.1/31.5 | 38.7/29.6 | 39.7/31.6 | 39.3/31.4 | 39.7/29.5 | 40.7/29.6 |
| | SRDE(10°) | 0.9% | 1.8% | 1.4% | 1.4% | 0.9% | 3.2% | 3.2% | 3.5% | 3.0% | 2.3% |
| | SRDE(20°) | 3.5% | 4.8% | 4.4% | 3.4% | 4.4% | 11.5% | 10.9% | 11.3% | 8.6% | 7.6% |
| | SRDE(30°) | 9.7% | 9.0% | 7.9% | 8.6% | 10.2% | 20.6% | 18.3% | 18.7% | 17.6% | 13.1% |
| **TL-SSC[18]** | $\phi$ error (mean/deviation)/° | 47.6/40.3 | 51.4/40.7 | 58.1/43.2 | 67.1/44.5 | 70.2/45.1 | 37.8/37.3 | 41.9/38.4 | 50.1/40.3 | 59.8/42.6 | 64.1/43.3 |
| | $\theta$ error (mean/deviation)/° | 19.0/11.4 | 19.0/11.4 | 19.0/11.5 | 19.1/11.6 | 19.1/11.7 | 18.9/11.5 | 18.8/11.3 | 18.7/11.3 | 18.9/11.5 | 18.9/11.5 |
| | SRDE(10°) | 4.8% | 5.5% | 5.5% | 3.5% | 4.8% | 3.9% | 4.9% | 3.0% | 3.0% | 3.7% |
| | SRDE(20°) | 18.3% | 15.3% | 12.5% | 7.9% | 7.2% | 27.0% | 23.3% | 16.0% | 9.7% | 9.7% |
| | SRDE(30°) | 36.9% | 31.9% | 25.6% | 18.0% | 16.2% | 43.9% | 40.2% | 33.0% | 23.6% | 21.3% |
| **LS-SVM[19]** | $\phi$ error (mean/deviation)/° | 46.0/39.9 | 57.5/43.7 | 70.2/48.5 | 58.9/43.0 | - | 26.4/31.4 | 27.6/29.6 | 52.7/45.2 | 60.9/46.7 | - |
| | $\theta$ error (mean/deviation)/° | 36.2/26.8 | 36.6/26.1 | 39.0/28.2 | 39.2/27.7 | - | 33.6/24.3 | 34.5/25.9 | 35.9/25.5 | 39.3/29.2 | - |
| | SRDE(10°) | 3.9% | 1.8% | 1.6% | 1.6% | - | 9.0% | 7.8% | 2.5% | 3.0% | - |
| | SRDE(20°) | 13.4% | 7.1% | 6.5% | 6.9% | - | 24.5% | 20.3% | 12.2% | 7.2% | - |
| | SRDE(30°) | 23.3% | 15.3% | 12.5% | 13.2% | - | 41.4% | 35.3% | 22.1% | 16.1% | - |

| SNR/dB | | 0 | | | | | 10 | | | | |
|---|---|---|---|---|---|---|---|---|---|---|---|
| $T_{60}$/ms | | 0 | 100 | 200 | 400 | 600 | 0 | 100 | 200 | 400 | 600 |
| **GCA (Proposed)** | $\phi$ error (mean/deviation)/° | 1.1/0.9 | 1.1/1.0 | 1.5/1.2 | 3.5/6.2 | 6.4/8.7 | 1.1/0.9 | 1.0/0.9 | 1.3/1.1 | 3.5/10.1 | 6.7/13.2 |
| | $\theta$ error (mean/deviation)/° | 0.9/0.7 | 0.9/0.7 | 1.1/0.9 | 2.3/3.9 | 4.0/4.8 | 0.9/0.7 | 0.8/0.7 | 1.0/0.8 | 1.8/2.9 | 3.8/6.3 |
| | SRDE(10°) | 100% | 100% | 100% | 93.5% | 77.8% | 100% | 100% | 100% | 94.5% | 81.7% |
| | SRDE(20°) | 100% | 100% | 100% | 99.0% | 92.1% | 100% | 100% | 100% | 98.8% | 91.0% |
| | SRDE(30°) | 100% | 100% | 100% | 99.5% | 97.2% | 100% | 100% | 100% | 99.1% | 94.5% |
| **TDE[14]** | $\phi$ error (mean/deviation)/° | 51.8/55.0 | 58.8/57.7 | 56.2/55.8 | 60.6/55.3 | 60.7/57.2 | 33.5/51.9 | 46.9/56.5 | 53.9/27.5 | 58.2/57.3 | 57.1/57.4 |
| | $\theta$ error (mean/deviation)/° | 31.5/26.8 | 33.8/27.9 | 35.1/28.9 | 36.7/29.6 | 35.9/27.2 | 24.0/23.7 | 27.6/25.0 | 31.0/27.5 | 31.9/26.4 | 33.9/27.4 |
| | SRDE(10°) | 11.3% | 6.9% | 7.2% | 4.9% | 4.1% | 26.1% | 16.1% | 12.5% | 9.4% | 8.1% |
| | SRDE(20°) | 24.5% | 20.8% | 20.8% | 17.3% | 15.2% | 35.5% | 28.2% | 25.4% | 20.6% | 19.9% |
| | SRDE(30°) | 37.9% | 33.3% | 33.2% | 27.0% | 26.5% | 54.9% | 47.4% | 42.9% | 35.6% | 36.5% |
| **TL-SSC[18]** | $\phi$ error (mean/deviation)/° | 29.2/33.9 | 34.0/36.2 | 42.7/39.5 | 53.3/41.1 | 59.2/42.2 | 20.2/28.6 | 22.8/32.3 | 30.5/36.0 | 44.3/39.8 | 52.7/41.8 |
| | $\theta$ error (mean/deviation)/° | 18.9/11.6 | 18.7/11.4 | 18.6/11.4 | 18.7/11.4 | 18.8/11.5 | 19.5/11.2 | 19.2/11.3 | 18.6/11.3 | 18.6/11.4 | 18.7/11.3 |
| | SRDE(10°) | 6.7% | 5.1% | 3.9% | 3.5% | 3.3% | 10.6% | 9.5% | 6.5% | 3.5% | 2.3% |
| | SRDE(20°) | 35.8% | 31.4% | 22.4% | 12.5% | 10.8% | 39.5% | 41.6% | 33.9% | 18.5% | 13.6% |
| | SRDE(30°) | 51.0% | 49.7% | 39.5% | 28.4% | 24.5% | 64.2% | 61.4% | 54.7% | 36.2% | 31.2% |
| **LS-SVM[19]** | $\phi$ error (mean/deviation)/° | 24.3/28.0 | 22.8/27.7 | 39.3/43.5 | 57.1/51.9 | - | 29.3/38.9 | 27.7/35.5 | 24.2/28.9 | 29.1/29.1 | - |
| | $\theta$ error (mean/deviation)/° | 34.2/25.2 | 33.1/24.5 | 36.6/28.8 | 39.2/29.0 | - | 35.4/25.6 | 31.7/24.1 | 35.4/26.1 | 36.4/27.7 | - |
| | SRDE(10°) | 12.2% | 10.4% | 7.2% | 3.0% | - | 9.4% | 11.1% | 10.6% | 5.3% | - |
| | SRDE(20°) | 24.2% | 26.3% | 18.9% | 11.1% | - | 21.2% | 25.9% | 24.7% | 21.2% | - |
| | SRDE(30°) | 37.0% | 40.7% | 34.7% | 23.1% | - | 38.0% | 42.2% | 37.6% | 35.5% | - |

TABLE III  (A) ROBUSTNESS VALIDATION FOR PROPOSED GCA WITH FIXED SNR AND VARYING $T_{60}$ AT 567 TEST POSITIONS

| SNR/dB | −10 | | | | | −5 | | | | | 0 | | | | | 10 | | | | |
|---|---|---|---|---|---|---|---|---|---|---|---|---|---|---|---|---|---|---|---|---|
| $T_{60}$/ms (train) | 200 | | | | | 200 | | | | | 200 | | | | | 200 | | | | |
| $T_{60}$/ms (localize) | 0 | 100 | 200 | 400 | 600 | 0 | 100 | 200 | 400 | 600 | 0 | 100 | 200 | 400 | 600 | 0 | 100 | 200 | 400 | 600 |
| SRDE(10°) | 67.2% | 64.9% | 84.3% | 44.3% | 35.1% | 74.4% | 76.9% | 99.5% | 76.4% | 66.7% | 69.0% | 76.4% | 100% | 83.8% | 79.5% | 81.1% | 86.1% | 100% | 94.2% | 91.9% |
| SRDE(20°) | 96.3% | 95.1% | 97.9% | 75.0% | 64.2% | 97.5% | 98.2% | 100% | 96.5% | 91.0% | 96.8% | 98.2% | 100% | 99.0% | 97.0% | 98.2% | 98.8% | 100% | 99.8% | 99.1% |
| SRDE(30°) | 99.5% | 98.9% | 99.8% | 82.5% | 77.1% | 99.1% | 99.5% | 100% | 99.5% | 93.8% | 100% | 100% | 100% | 100% | 98.9% | 100% | 100% | 100% | 100% | 99.5% |

(B) ROBUSTNESS VALIDATION FOR PROPOSED GCA WITH FIXED $T_{60}$ AND VARYING SNR AT 567 TEST POSITIONS

| $T_{60}$/ms | 0 | | | | 100 | | | | 200 | | | | 400 | | | | 600 | | | |
|---|---|---|---|---|---|---|---|---|---|---|---|---|---|---|---|---|---|---|---|---|
| SNR/dB (train) | 0 | | | | 0 | | | | 0 | | | | 0 | | | | 0 | | | |
| SNR/dB (localize) | −10 | −5 | 0 | 10 | −10 | −5 | 0 | 10 | −10 | −5 | 0 | 10 | −10 | −5 | 0 | 10 | −10 | −5 | 0 | 10 |
| SRDE(10°) | 76.9% | 86.8% | 100% | 96.0% | 76.2% | 95.1% | 100% | 83.4% | 64.4% | 92.4% | 100% | 68.8% | 34.2% | 72.8% | 93.5% | 72.7% | 25.1% | 58.9% | 77.8% | 68.4% |
| SRDE(20°) | 87.1% | 98.4% | 100% | 99.8% | 85.8% | 100% | 100% | 98.8% | 86.8% | 99.8% | 100% | 96.8% | 59.1% | 90.8% | 99.0% | 95.6% | 47.1% | 80.1% | 92.1% | 92.2% |
| SRDE(30°) | 94.3% | 99.3% | 100% | 100% | 93.3% | 100% | 100% | 99.5% | 94.9% | 99.8% | 100% | 99.8% | 70.1% | 94.7% | 99.5% | 99.4% | 61.0% | 88.7% | 97.2% | 96.6% |

(C) RESULTS OF PROPOSED GCA BY SOUND SOURCES AT DIFFERENT FREQUENCIES WITH 567 TEST POSITIONS

| Source | | Machinery sound | | | | | | | | | Telephone ring | | | | | | | | | |
|---|---|---|---|---|---|---|---|---|---|---|---|---|---|---|---|---|---|---|---|---|
| SNR/dB | | −10 | | | | | 10 | | | | | −10 | | | | | 10 | | | | |
| $T_{60}$/ms | | 0 | 100 | 200 | 400 | 600 | 0 | 100 | 200 | 400 | 600 | 0 | 100 | 200 | 400 | 600 | 0 | 100 | 200 | 400 | 600 |
| **GCA (Proposed)** | SRDE(10°) | 98.9% | 98.8% | 87.7% | 50.8% | 26.6% | 99.3% | 99.8% | 99.1% | 84.1% | 69.7% | 100% | 98.4% | 71.6% | 27.7% | 12.7% | 97.4% | 97.7% | 90.8% | 47.6% | 28.2% |
| | SRDE(20°) | 100% | 100% | 98.9% | 81.8% | 55.9% | 100% | 100% | 100% | 94.0% | 84.0% | 100% | 100% | 93.3% | 58.9% | 37.0% | 100% | 100% | 98.2% | 77.3% | 56.4% |
| | SRDE(30°) | 100% | 100% | 99.5% | 91.4% | 71.6% | 100% | 100% | 100% | 96.8% | 90.1% | 100% | 100% | 98.1% | 76.4% | 52.9% | 100% | 100% | 99.3% | 88.7% | 72.0% |

TABLE IV  RESULTS OF THE FOUR ALGORITHMS AT 378 TEST POSITIONS

| SNR/dB | | −10 | | | | | −5 | | | | | 0 | | | | | 10 | | | | |
|---|---|---|---|---|---|---|---|---|---|---|---|---|---|---|---|---|---|---|---|---|---|
| $T_{60}$/ms | | 0 | 100 | 200 | 400 | 600 | 0 | 100 | 200 | 400 | 600 | 0 | 100 | 200 | 400 | 600 | 0 | 100 | 200 | 400 | 600 |
| **GCA (Proposed)** | SRDE(10°) | 100% | 99.5% | 82.0% | 48.9% | 31.7% | 100% | 100% | 99.2% | 76.5% | 57.9% | 100% | 100% | 100% | 91.8% | 72.5% | 100% | 100% | 100% | 93.4% | 80.1% |
| | SRDE(20°) | 100% | 100% | 96.8% | 73.8% | 57.9% | 100% | 100% | 100% | 94.4% | 79.4% | 100% | 100% | 100% | 98.4% | 90.0% | 100% | 100% | 100% | 98.9% | 91.3% |
| | SRDE(30°) | 100% | 100% | 99.7% | 87.3% | 70.1% | 100% | 100% | 100% | 97.6% | 87.8% | 100% | 100% | 100% | 99.2% | 96.3% | 100% | 100% | 100% | 99.2% | 93.9% |
| **TDE[14]** | SRDE(10°) | 0.8% | 2.1% | 1.3% | 1.6% | 1.1% | 3.2% | 3.2% | 3.4% | 2.6% | 1.9% | 11.6% | 7.9% | 7.1% | 5.0% | 5.0% | 27.5% | 15.1% | 15.1% | 11.6% | 9.8% |
| | SRDE(20°) | 3.7% | 5.0% | 4.0% | 3.7% | 5.6% | 11.9% | 10.6% | 11.9% | 8.5% | 8.2% | 25.9% | 20.6% | 20.4% | 18.8% | 16.9% | 36.5% | 27.8% | 28.3% | 24.1% | 19.8% |
| | SRDE(30°) | 10.1% | 10.1% | 7.4% | 9.8% | 10.6% | 20.4% | 16.1% | 20.1% | 17.2% | 13.2% | 40.2% | 32.5% | 32.5% | 28.0% | 27.5% | 55.0% | 48.9% | 44.7% | 38.7% | 37.0% |
| **TL-SSC [18]** | SRDE(10°) | 4.5% | 5.6% | 5.6% | 4.0% | 5.0% | 3.4% | 4.2% | 2.4% | 3.2% | 3.4% | 6.1% | 3.7% | 3.7% | 3.2% | 3.2% | 7.1% | 6.6% | 4.5% | 3.1% | 2.6% |
| | SRDE(20°) | 16.7% | 13.2% | 12.4% | 7.4% | 7.4% | 24.9% | 21.4% | 13.5% | 9.0% | 9.0% | 33.9% | 27.3% | 20.4% | 11.4% | 8.5% | 37.0% | 39.2% | 29.9% | 16.4% | 14.1% |
| | SRDE(30°) | 35.4% | 30.4% | 24.9% | 18.0% | 15.9% | 42.6% | 38.6% | 31.0% | 23.0% | 19.8% | 48.2% | 46.6% | 36.5% | 26.2% | 22.5% | 62.8% | 58.0% | 50.7% | 33.8% | 28.9% |
| **LS-SVM [19]** | SRDE(10°) | 4.5% | 2.4% | 1.1% | 1.1% | - | 10.3% | 8.7% | 2.1% | 2.4% | - | 12.4% | 11.4% | 7.4% | 3.2% | - | 10.6% | 10.9% | 12.4% | 5.3% | - |
| | SRDE(20°) | 14.6% | 8.5% | 6.1% | 6.6% | - | 25.9% | 21.7% | 13.0% | 5.8% | - | 22.2% | 24.6% | 20.4% | 12.4% | - | 23.0% | 27.8% | 26.7% | 22.8% | - |
| | SRDE(30°) | 25.7% | 16.7% | 11.9% | 11.6% | - | 44.7% | 38.1% | 22.5% | 14.6% | - | 40.0% | 39.4% | 37.3% | 23.3% | - | 42.2% | 41.6% | 40.2% | 37.0% | - |

TABLE V  THE COMPUTATIONAL COMPLEXITY BY CPU TIME (S) AND REAL TIME (S) AT 567 TEST POSITIONS

| | | $T_{60}$=0ms | | $T_{60}$=100ms | | $T_{60}$=200ms | | $T_{60}$=400ms | | $T_{60}$=600ms | |
|---|---|---|---|---|---|---|---|---|---|---|---|
| | | CPU time | Real time | CPU time | Real time | CPU time | Real time | CPU time | Real time | CPU time | Real time |
| GCA (Proposed) | Offline training | 2747.7 | 1188.8 | 3950.4 | 2370.4 | 6012.1 | 4618.5 | 12865.6 | 11586.4 | 31067.4 | 29298.3 |
| | Online localization | 24.5 | 9.6 | 21.7 | 9.4 | 19.3 | 9.2 | 17.2 | 11.8 | 20.1 | 9.8 |
| TDE[14] | Online localization | 32911.0 | 30149.3 | 37023.8 | 34708.9 | 42052.5 | 39121.1 | 55222.1 | 51748.0 | 58945.9 | 55107.0 |
| TL-SSC[18] | Offline LUT computing | 906.9 | 653.1 | 894.7 | 643.6 | 898.4 | 643.4 | 941.9 | 717.3 | 927.1 | 716.2 |
| | Online localization | 743.5 | 583.4 | 741.9 | 576.9 | 755.4 | 580.5 | 843.9 | 699.3 | 810.4 | 686.9 |
| LS-SVM[19] | Offline training | 374300.9 | 28594.9 | 367334.3 | 29065.0 | 380892.9 | 32471.0 | 391769.3 | 37791.7 | - | - |
| | Online localization | 21.4 | 9.6 | 18.2 | 8.5 | 18.5 | 8.8 | 19.9 | 8.4 | - | - |

TABLE VI  TRADEOFF BETWEEN LOCALIZATION ACCURACY AND COMPUTATIONAL COMPLEXITY AT 567 TEST POSITIONS

| $K$ | 512 | 4096 | 32768 |
|---|---|---|---|
| SRDE(10°) | 86.9% | 99.6% | 99.8% |
| SRDE(20°) | 99.8% | 100% | 100% |
| SRDE(30°) | 100% | 100% | 100% |

| $K$ | 512 | | 4096 | | 32768 | |
|---|---|---|---|---|---|---|
| | CPU time(s) | Real time(s) | CPU time(s) | Real time(s) | CPU time(s) | Real time(s) |
| Offline training | 1891.8 | 252.5 | 3900.1 | 2255.5 | 140462.3 | 21071.5 |
| Online localization | 3.8 | 0.6 | 23.1 | 9.8 | 3595.8 | 1815.5 |

TL-SSC, GCA spends more CPU time. However, this computational overhead is acceptable considering the significant improvements of 82.6%, 74.1%, and 60.3% by GCA over TL-SSC for SRDE(10°), SRDE(20°), and SRDE(30°) respectively.

*Tradeoff strategy*

The number of clusters $K$ is proportional to computational complexity. When $K$ is small, although computational complexity is inexpensive, the features become inconsistent, resulting in degradation of localization accuracy. In contrast, if $K$ is large, although the features become consistent, computational complexity becomes expensive or even unaffordable. Therefore, the quantity of space cluster division should be determined by making a tradeoff between localization accuracy and computational complexity.

To illustrate this kind of tradeoff, we conduct experiments by varying $K$, in the environment where $T_{60}$ = 100ms and SNR = –10dB. $K$ is set to be 512, 4096, and 32768, corresponding to cluster volume of 0.5m×0.5m×0.5m, 0.25m×0.25m×0.25m, and 0.125m×0.125m×0.125m. The results are summarized in Table VI. From the results, it can be observed that the complexity increases with the growth of $K$. When $K$=4096, compared to the case of $K$=512, SRDE(10°) is significantly improved by 12.7%, achieving to 99.6%, although the cost is some complexity increase. When $K$ increases from 4096 to 32768, SRDE(10°) can hardly be improved further, however, the complexity becomes very expensive. Therefore, we divide the room into 4096 clusters.

## V. CONCLUSION

In this paper, we address the problem of SSL in the challenging high reverberation and low SNR environments by proposing a novel machine learning based algorithm GCA. With GCC feature, the proposed GCA transforms the SSL problem into a likelihood based nonlinear classification problem by utilizing PNN, which is especially suitable for multiclass classification problem. In order to overcome the misclassification and estimate DOA more accurately, we propose WLDM in GCA. The experimental results have shown that GCA achieves more accurate DOA estimation. The average of mean values of azimuth angle estimation errors and elevation angle estimation errors of GCA are only 4.6° and 3.1° respectively. Compared with three recently published algorithms, GCA improves the best performances of average SRDE(10°), SRDE(20°) and SRDE(30°) by 81.1%, 74.1% and 60.3% respectively. In addition, GCA performs robustly in different acoustic environments. This validates that the proposed GCA can localize very effectively for the applications when physical site acoustic features can be accessed before the localization stage. This data driven training method is especially suitable for the industry environments which are too complex to be modeled.

REFERENCES

[1] H. Guo, K. S. Low, and H. A. Nguyen, "Optimizing the localization of a wireless sensor network in real time based on a low-cost microcontroller," *IEEE Trans. Ind. Electron.*, vol. 58, no. 3, pp. 741–749, Mar. 2011.
[2] B. Wang, S. Zhou, W. Liu, and Y. Mo, "Indoor localization based on curve fitting and location search using received signal strength," *IEEE Trans. Ind. Electron.*, vol. 62, no. 1, pp. 572–582, Jan. 2015.
[3] F. Deng et al., "Energy-based sound source localization with low power consumption in wireless sensor networks," *IEEE Trans. Ind. Electron.*, vol. 64, no. 6, pp. 4894–4902, Jun. 2017.
[4] P. Yang and W. Wu, "Efficient particle filter localization algorithm in dense passive RFID tags environment," *IEEE Trans. Ind. Electron.*, vol. 61, no. 10, pp. 5641–5651, Oct. 2014.
[5] J. Kim and W. Chung, "Localization of a mobile robot using a laser range finder in a glass-walled environment," *IEEE Trans. Ind. Electron.*, vol. 63, no. 6, pp. 3616–3627, Jun. 2016.
[6] H. Song, W. Choi, and H. Kim, "Robust vision-based relative-localization approach using an RGB-depth camera and LiDAR sensor fusion," *IEEE Trans. Ind. Electron.*, vol. 63, no. 6, pp. 3725–3736, Jun. 2016.
[7] J. Wang, Q. Gao, Y. Yu, H. Wang, and M. Jin, "Toward robust indoor localization based on Bayesian filter using chirp-spread-spectrum ranging," *IEEE Trans. Ind. Electron.*, vol. 59, no. 3, pp. 1622–1629, Mar. 2012.
[8] J. Wang et al., "Transferring compressive-sensing-based device-free localization across target diversity," *IEEE Trans. Ind. Electron.*, vol. 62, no. 4, pp. 2397–2409, Apr. 2015.
[9] J. Chen, J. Benesty, and Y. Huang, "Time delay estimation in room acoustic environments: An overview," *EURASIP J. Appl. Signal Process.*, vol. 2006, pp. 1–19, Jan. 2006.
[10] S. Argentieri, P. Danès, and P. Souères, "A survey on sound source localization in robotics: From binaural to array processing methods," *Comput. Speech Lang.*, vol. 34, no. 1, pp. 87–112, Nov. 2015.
[11] H. He, L. Wu, J. Lu, X. Qiu, and J. Chen, "Time difference of arrival estimation exploiting multichannel spatio-temporal prediction," *IEEE Trans. Audio Speech Lang. Process.*, vol. 21, no. 3, pp. 463–475, Mar. 2013.
[12] D. Pavlidi, A. Griffin, M. Puigt, and A. Mouchtaris, "Real-time multiple sound source localization and counting using a circular microphone array," *IEEE Trans. Audio Speech Lang. Process.*, vol. 21, no. 10, pp. 2193–2206, Oct. 2013.
[13] A. Canclini, E. Antonacci, A. Sarti, and S. Tubaro, "Acoustic source localization with distributed asynchronous microphone networks," *IEEE Trans. Audio Speech Lang. Process.*, vol. 21, no. 2, pp. 439–443, Feb. 2013.
[14] X. Alameda-Pineda and R. Horaud, "A geometric approach to sound source localization from time-delay estimates," *IEEE/ACM Trans. Audio Speech Lang. Process.*, vol. 22, no. 6, pp. 1082–1095, Jun. 2014.
[15] J. Velasco, C. J. Martn-Arguedas, J. Macias-Guarasa, D. Pizarro, and M. Mazo, "Proposal and validation of an analytical generative model of

SRP-PHAT power maps in reverberant scenarios," *Signal Process.*, vol. 119, pp. 209–228, Feb. 2016.
[16] B. Mungamuru and P. Aarabi, "Enhanced sound localization," *IEEE Trans. Syst. Man Cybern. B Cybern.*, vol. 34, no. 3, pp. 1526–1540, Jun. 2004.
[17] J. Dmochowski, J. Benesty, and S. Affes, "A generalized steered response power method for computationally viable source localization," *IEEE Trans. Audio Speech Lang. Process.*, vol. 15, no. 8, pp. 2510–2526, Nov. 2007.
[18] D. Yook, T. Lee, and Y. Cho, "Fast sound source localization using two-level search space clustering," *IEEE Trans. Cybern.*, vol. 46, no. 1, pp. 20–26, Jan. 2016.
[19] H. Chen and W. Ser, "Acoustic source localization Using LS-SVMs without calibration of microphone arrays," in *Proc. IEEE Int. Symp. Circuits Syst.*, Taipei, Taiwan, May 2009, pp. 1863–1866.
[20] X. Xiao et al., "A learning-based approach to direction of arrival estimation in noisy and reverberant environments," in *Proc. IEEE Int. Conf. Acoust. Speech Signal Process.*, Brisbane, Australia, Apr. 2015, pp. 76–80.
[21] X. Li and H. Liu, "Sound source localization for HRI using FOC-based time difference feature and spatial grid matching," *IEEE Trans. Cybern.*, vol. 43, no. 4, pp. 1199–1212, Aug. 2013.
[22] B. Laufer-Goldshtein, R. Talmon, and S. Gannot, "Semi-supervised sound source localization based on manifold regularization," *IEEE/ACM Trans. Audio Speech Lang. Process.*, vol. 24, no. 8, pp. 1393–1407, Aug. 2016.
[23] D. F. Specht, "Probabilistic neural networks," *Neural Netw.*, vol. 3, no. 1, pp. 109–118, Jan. 1990.
[24] E. Lehmann and A. Johansson, "Diffuse reverberation model for efficient image-source simulation of room impulse responses," *IEEE Trans. Audio Speech Lang. Process.*, vol. 18, no. 6, pp. 1429–1439, Aug. 2010.
[25] Y. Hu and P. Loizou. NOIZEUS database. [Online]. Available: http://ecs.utdallas.edu/loizou/speech/noizeus.
[26] X. Alameda-Pineda and R. Horaud. The gtde MATLAB toolbox. [Online]. Available: https://team.inria.fr/perception/research/geometric-sound-source-localization.
[27] T. Lee. TLSSC code. [Online]. Available: https://github.com/LeeTaewoo/fast_sound_source_localization_using_TLSSC.
[28] E. Lehmann. Fast ISM code. [Online]. Available: http://www.eric-lehmann.com.
[29] FindSounds database. [Online]. Available: http://www.findsounds.com.

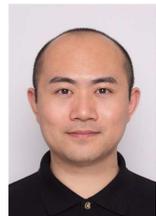
**Yingxiang Sun** (S'16) received his B.Eng. degree in Electronic and Information Engineering from Xidian University, Xi'an, China, in 2009. He received his M.Eng. degree in Electromagnetic Field and Microwave Technology from the 54th Research Institute of China Electronics Technology Group Corporation, Shijiazhuang, China, in 2012. Currently, he is working towards the Ph.D. degree at the Pillar of Engineering Product Development, Singapore University of Technology and Design, Singapore. His current research interests include digital signal processing and machine learning.

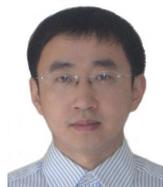
**Jiajia Chen** received his B. Eng. (Hons) and Ph.D. from Nanyang Technological University, Singapore, in 2004 and 2010, respectively. Since April 2012, he has been with Singapore University of Technology and Design, where he is currently a Senior Lecturer. His research interest includes computational transformations of low-complexity digital filters, image fusion and audio signal processing. Dr. Chen served as Web Chair of *Asia-Pacific Computer Systems Architecture Conference* 2005, Technical Program Committee member of *European Signal Processing Conference* 2014 and *The Third IEEE International Conference on Multimedia Big Data* 2017, and Associate Editor of *Springer EURASIP Journal on Embedded Systems* since 2016.

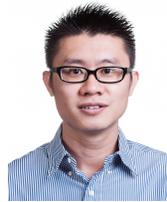
**Chau Yuen** (S'02–M'08–SM'12) received the B.Eng. and Ph.D. degrees from Nanyang Technological University, Singapore, in 2000 and 2004, respectively. In 2005, he was a Post-Doctoral Fellow with Lucent Technologies Bell Labs, Murray Hill, NJ, USA. In 2008, he was a Visiting Assistant Professor with Hong Kong Polytechnic University, Hong Kong. From 2006 to 2010, he was a Senior Research Engineer with the Institute for Infocomm Research, Singapore, where he was involved in an industrial project developing an 802.11n wireless local area network system and actively participated in the third generation Partnership Project Long-Term Evolution (LTE) and LTE-A standardization. In 2010, he joined the Singapore University of Technology and Design, Singapore, as an Assistant Professor. He has authored over 300 research papers in international journals or conferences. He holds two U.S. patents. He received the IEEE Asia-Pacific Outstanding Young Researcher Award in 2012. He serves as an Editor of the *IEEE TRANSACTIONS ON COMMUNICATIONS* and *IEEE TRANSACTIONS ON VEHICULAR*.

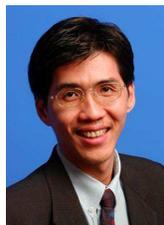
**Susanto Rahardja** (F'11) received the B.Eng. degree from National University of Singapore in 1991, the M.Eng. and Ph.D. degrees all in Electronic Engineering from Nanyang Technological University, Singapore, in 1993 and 1997 respectively. He is currently a Chair Professor at the Northwestern Polytechnical University (NPU) under the Thousand Talent Plan of People's Republic of China. His research interests are in multimedia, signal processing, wireless communications, discrete transforms and signal processing algorithms, implementation and optimization. Dr Rahardja was the recipients of numerous awards, including the IEE Hartree Premium Award, the Tan Kah Kee Young Inventors' Open Category Gold award, the Singapore National Technology Award, A*STAR Most Inspiring Mentor Award, Finalist of the 2010 World Technology & Summit Award, the Nokia Foundation Visiting Professor Award and the ACM Recognition of Service Award.